\documentclass[conference]{IEEEtran}
\IEEEoverridecommandlockouts
\usepackage{cite}
\usepackage{amsmath,amssymb,amsfonts}
\usepackage{algorithmic}
\usepackage{graphicx}
\usepackage{textcomp}
\usepackage{xcolor}
\usepackage{url}
\def\BibTeX{{\rm B\kern-.05em{\sc i\kern-.025em b}\kern-.08em
    T\kern-.1667em\lower.7ex\hbox{E}\kern-.125emX}}
\begin{document}

\title{Probabilistic Agents in Deterministic Audits: Evaluating Multi-Agent Systems for Automated Audits Based on the German IT-Grundschutz}

\author{\IEEEauthorblockN{Lea Muth}
\IEEEauthorblockA{\textit{Department of Mathematics and Computer Science} \\
\textit{Freie Universität Berlin}\\
Berlin, Germany \\
Lea.Muth@fu-berlin.de}
\and
\IEEEauthorblockN{Marian Margraf}
\IEEEauthorblockA{\textit{Department of Mathematics and Computer Science} \\
\textit{Freie Universität Berlin}\\
Berlin, Germany \\
Marian.Margraf@fu-berlin.de}
 
}

\maketitle

\begingroup
\renewcommand\thefootnote{}
\footnotetext{\textcopyright~2026 IEEE. Personal use of this material is permitted. Permission from IEEE must be obtained for all other uses, in any current or future media, including reprinting/republishing this material for advertising or promotional purposes, creating new collective works, for resale or redistribution to servers or lists, or reuse of any copyrighted component of this work in other works.}
\endgroup

\begin{abstract}
The NIS-2 Directive mandates robust Risk Management from thousands of small and medium enterprises. To ensure compliance, companies rely on established standards such as the German IT-Grundschutz (IT-GS) of the Federal Office for Information Security. However, IT-GS certification is resource-intensive and requires a high level of manual effort for documentation, validation, and revision, making scalable implementation difficult and expensive. 

Building upon our previous conceptual framework, this paper presents the technical implementation and empirical evaluation of a Multi-Agent System (MAS) architecture combined with Hybrid Retrieval Augmented Generation (HybridRAG) for the partial automation of IT-GS certification. We introduce two novel technical contributions to the MAS architecture to enforce the compliance rigor. The Hypothesis-Verification Loop in the Structural Analysis (SA) phase that cross-references agent-inferred dependencies against the Knowledge Graph to reduce hallucinations, and a Decoupled Reasoning Pipeline that separates agent-driven semantic extraction from the deterministic protection need inheritance.
We utilize the BSI's "RecPlast GmbH" case study as a human expert-generated reference data set for end-to-end evaluation of the architecture and to quantify Precision, Recall, and F1-scores. The performance of the system is investigated across the phases of SA, Protection Needs Assessment (PNA), Modeling, and IT-GS Check.

The empirical results reveal noticeable differences throughout the different steps of IT-GS. While the MAS demonstrates high efficacy in semantic tasks (SA and Modeling), significantly reducing manual effort through automated information extraction, quantitative results reveal limitations in logical reasoning phases (PNA and IT-GS Check) as the probabilistic nature of current LLMs struggles to meet the deterministic rigor required by IT-GS. 
\end{abstract}


\section{Introduction}
Due to the Network and Information Security Directive 2 (NIS-2)~\cite{NIS2}, the EU expands the circle of companies legally required to have verifiable state-of-the-art risk management from a few critical infrastructure operators to thousands of small and medium Enterprises (SMEs), thus, creating significant compliance pressure for SMEs~\cite{NIS2}.
To demonstrate compliance, companies rely on established standards for implementation of Information Security Management System (ISMS). Internationally, it is ISO/IEC 27001~\cite{iso_27001} and in Germany it is the Federal Office for Information Security's (BSI) IT-Grundschutz (IT-GS) methodology~\cite{grundschutz_standards}. IT-GS is considered the gold standard for the state-of-the-art due diligence required according to NIS-2~\cite{BSI_NIS2_2025}.
Scalability of IT security audits remains an ongoing challenge, especially for SMEs, given increasing regulatory requirements, high certification costs, an ongoing shortage of skilled workers, and continuously changing IT infrastructures~\cite{BSI_NIS2_2025, enisa}. 
Since IT-GS relies to a large extent on the processing of extensive text documentation, Large Language Models (LLMs) appear as clear candidates for automation. However, their inherent probabilistic nature conflicts with the strict demands of a compliance audit, which requires deterministic evidence and verifiable facts.

Although our previous work~\cite{former_paper} proposed an architecture of a supporting Multi-Agent System (MAS) based on a HybridRAG~\cite{hybridRAG, graphRAG_OG} for semi-automated certification based on the German IT-GS, the practical implementation of the system in a strict audit context remained to be done. 
This paper provides the first empirical end-to-end evaluation of the MAS pipeline against the BSI's "RecPlast GmbH" case study~\cite{recplast}. Unlike real-world corporate data, where confidentiality prevents the availability of a transparent data set, RecPlast serves as a ground truth for a complete IT-GS implementation. While it provides an idealized structural foundation, it is an expert-generated data set for verifying the accuracy of our agent-based framework and quantifying metrics such as recall and precision.
The applicability of HybridRAG for enforcing audit compliance and mitigating the risk of hallucinations by enforcing consistency through deterministic validation in the Structural Analysis (SA) is demonstrated. While our findings confirm that full automation remains elusive, they highlight the shift within the auditor's workflow. The MAS can move the auditor's role from manual data collection to the time-saving validation of AI-generated content in the SA and Modeling. 
\section{Background}
\subsection{The IT-Grundschutz Certification Process}
\begin{figure}[htbp]
\centerline{\includegraphics[width=0.268\textwidth]{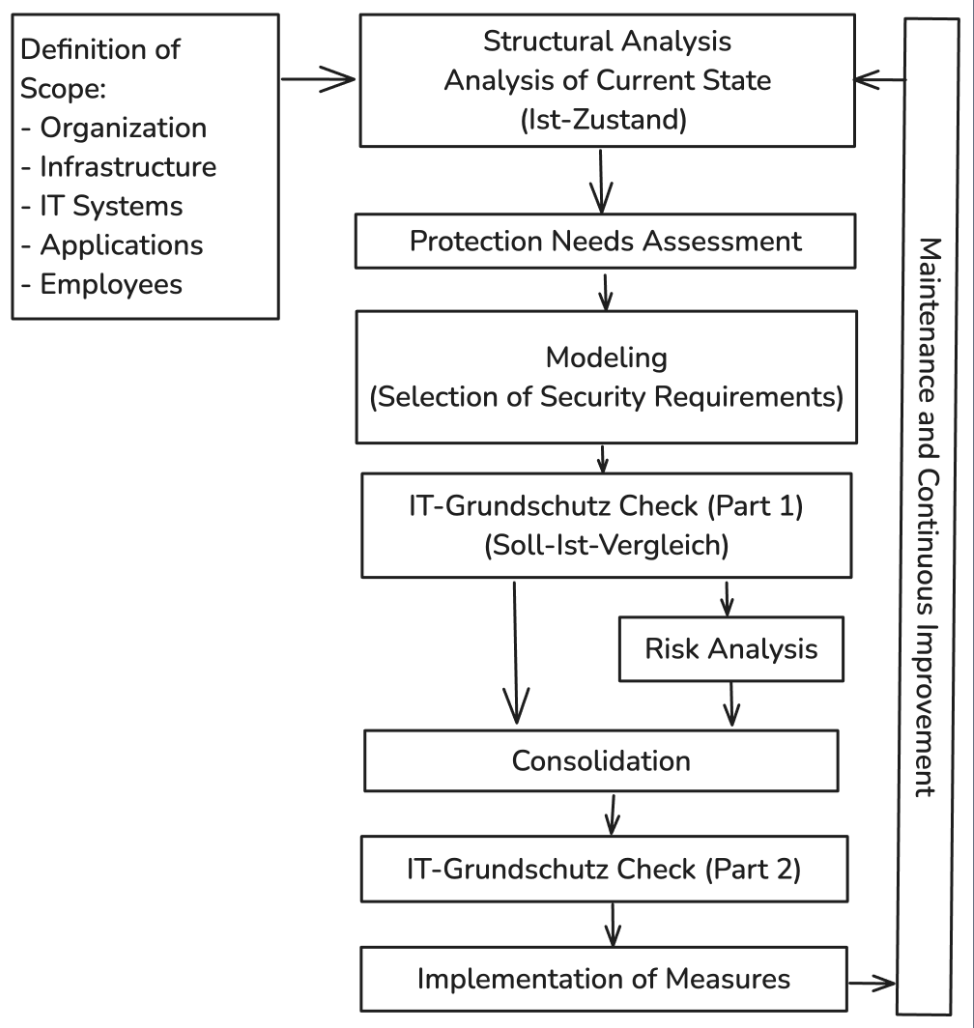}}
\caption{Process of the IT-Grundschutz Certification. The diagram shows the nine-step process for carrying out certification under ISO 27001 based on IT-Grundschutz.}
\label{fig}
\end{figure}

The IT-GS is a comprehensive methodology developed by the German BSI to provide organizations with a systematic approach to information security. By achieving compliance, organizations demonstrate an established security level, which is verified through the IT-GS certification process ("ISO 27001 certificate based on IT-GS"), as illustrated in Fig.~\ref{fig}. 
The methodology is based on four BSI-Standards~\cite{grundschutz_standards} specifying the general requirements and structure of an ISMS, elaborating Risk Analysis, and Business Continuity Management.
Within the \textbf{Definition of the Scope}, the boundaries of the ISMS are set, determining which assets, processes, departments, and activities are covered by the ISMS. The \textbf{SA} extracts all components of the information network and their interactions within the defined scope. Being part of the SA, the \textbf{Dependency Analysis} (DA) captures the technical and organizational dependencies within a company's infrastructure as they are essential for understanding how security requirements spread throughout the company and determining their vulnerability propagation.
The SA forms the foundation of the certification process. Errors occurring here are critical because they result in assets being excluded from the security concept, e.g., a server not listed in the SA does not exist in the Security Concept and is prone to not being patched or monitored. An overlooked dependency in the DA is a critical failure, as it breaks the inheritance chain of Protection Needs (PNs), leaving high-risk assets potentially under-protected and invisible within the security concept.

The \textbf{Protection Needs Assessment} then determines the confidentiality, integrity and availability (CIA) requirements for each entity. PNs are categorized as "Normal", "High", or "Very High". The CIA security objectives are at the center of damage analysis and must be independently evaluated.
After initial determination, the PNs are systematically inherited following four deterministic principles: the Maximum Principle, Cumulative Effect, Distribution Effect, and DA.
Errors in determining the level of protection required destroy either security (underestimation) or cost-effectiveness (overestimation). The rules of inheritance create a complex dependency graph where changes to the PNs of a single entity can trigger system-wide propagation effects. Consequently, the correct application of these rules requires a continuous re-evaluation of the entire inheritance chain, as required by NIS-2.

To provide a comprehensive and standardized solution for various infrastructures, the \textbf{Modeling} replicates the information network using the standardized modules from the IT-GS Compendium~\cite{grundschutz_kompendium}. Each module (e.g., "web server") lists typical threats and specific security requirements that must be implemented to achieve a recognized level of protection ("Basic", "Standard", and "Enhanced Protection").
Technical components often require being mapped to multiple modules, e.g., a web server under Unix is mapped onto the modules general server, server under Unix, and web server to cover all security requirements of the component. 
An error in modeling creates a blind spot as it can lead to hundreds of concrete security requirements missing. In the event of a misalignment, requirements are established that are not actually relevant. 

The \textbf{IT-GS Check} covers +7,500 individual requirements for different security levels (basic, standard, and enhanced protection) and represents the actual degree of implementation within the company. Due to their granularity, they can be broken down into sub-requirements, which is particularly useful when measuring implementation progress. 
The basic requirements define the minimum level of security that every organization must implement. They address elementary risks, e.g., not storing passwords in plain text.  
Standard requirements define the state-of-the-art in terms of IT security and address typical threat scenarios, e.g., activating logging and implementing backup concepts.
The requirements for enhanced protection apply to critical systems and define additional measures for components whose compromise would threaten the existence of the company, e.g., encryption in idle mode and dedicated security monitoring systems.
Errors in IT-GS checks mean that known security vulnerabilities are incorrectly marked as resolved.

For components with increased PNs ("High" and "Very High"), for which no suitable modules exist, or that deviate from the standard assumptions of IT-GS due to special operating conditions, an explicit \textbf{Risk Analysis} (RA) is required to identify additional security requirements. During \textbf{Consolidation} measures are reviewed for conflicts and overlap to ensure coherent security measures without contradictory or redundant elements. Subsequently, the \textbf{Implementation of the measures} addresses all pending requirements through a realization plan. This plan includes prioritization based on urgency and effectiveness, binding deadlines, and a clear assignment of responsibilities. Finally, \textbf{Maintenance and Improvement} ensure the sustainability of the ISMS, requiring that the system is continuously maintained, monitored, and updated even after the certification process is completed.
\section{Literature Review}
The automation of IT security audits is a well-established area of research. Earlier approaches on automated compliance by~\cite{old2}, successfully mapped security standards such as ISO 27002 to formal ontologies to allow deterministic conclusions about threats, vulnerabilities, and controls. However, these semantic approaches were not scaling, as translating informal, unstructured control descriptions and organizational documents into formal, machine-readable rules requires considerable manual effort. Furthermore, as standards evolve, maintaining these rigid ontologies becomes economically unsustainable. Our work builds on the logical rigor of such ontological frameworks but overcomes the hurdle of manual modeling. Instead of human experts translating documents into formal logic, an agent autonomously extracts knowledge from raw source text and structures it into a Knowledge Graph (KG).

With the appearance of LLMs, the focus shifted to the direct processing of unstructured texts. The work of ~\cite{pilot_iso_27001} demonstrates the potential of LLMs for the planning phase of ISO 27001 audits based on prompt engineering and the comparison of several LLMs (Sonnet 3.5, GPT-4o, and Llama 3.1). In their work, the authors highlight clear limitations in scalability and the linking of complex information, focusing on the planning phase. They find that LLMs had difficulty correlating information across multiple documents, establishing linkages, and processing large amounts of data. This scalability issue led to the need for manual preparation and simplification of the data. Our approach addresses precisely this limitation by using a HybridRAG architecture in the context of BSI IT-GS, since a HybridRAG addresses this weakness of pure LLM prompts, since the graph makes the logical links explicit. To validate their results, the authors used a fictional company "HelloPak" to conduct their experiments. The evaluation of the results for "HelloPak" was not automated but was carried out manually by the authors themselves. Since there is very little ground truth for such subjective compliance tasks, the researchers relied on a combination of quantitative evaluation and qualitative analysis. Due to the high confidentiality of real audit data, the use of synthetic case studies is an accepted scientific practice in this field of research. Our work builds on this established practice, but benefits from a crucial qualitative enhancement as the data set for "RecPlast GmbH" was designed by the BSI and corresponds to a data set verified by experts.

In order to compensate for the limitations of pure LLM approaches, HybridRAG architectures that combine LLMs with KG are increasingly established. The authors of \cite{CCKG4HLRoV} present a framework that Common Vulnerabilities and Exposures (CVE) reports into KGs and enriches them with external databases (DBpedia) to allow for complex reasoning about threat scenarios. They demonstrate that providing an LLM with access to a HybridRAG significantly improves the quality of responses compared to LLM-only approaches. Our system uses a similar hybrid architecture, but shifts from threat analysis to compliance automation. While \cite{CCKG4HLRoV} rely on probabilistic relationships and external world knowledge, the BSI's IT-GS requires deterministic logic. Therefore, our solution replaces the open enrichment with strict rules (BSI Compendium) and focuses on the precise mapping of dependency chains for protection requirement inheritance, instead of causal attack paths.

Another application of HybridRAG in cybersecurity is presented by ~\cite{CyKGRAG}. Following a similar approach to our framework, they address the hallucination susceptibility of LLMs by integrating domain-specific cybersecurity KGs (CyKG). Their methodology combines vector embeddings for similarity searches with structured graph queries (SPARQL/Cypher) to compensate for the context loss of traditional RAG approaches. The key difference lies in the use case. While CyKG-RAG specializes in evaluating local security logs and dynamic threat data (e.g., CVE, MITRE ATT\&CK) for operational attack detection, our approach transfers the architecture to administrative compliance testing. In contrast to the exploratory nature of threat analysis, we use HybridRAG in combination with agents to build a verification system. The KG serves not only as a knowledge base, but also as a logical framework that enforces compliance with strict BSI regulations, thus ensuring the reliability required for audits, which pure LLM approaches cannot offer.
\section{Theoretical Framework: MAS Architecture and Experimental Setup}
To address the dual challenge of strict regulations and economic scalability for SMEs, we propose a decoupled architecture that separates knowledge management from audit execution. This separation allows us to apply more resource-intensive LLMs for the one-time creation of the semantic foundation in the form of the KG while enabling cost-efficient LLMs for the repetitive operational audit cycle.
The presented work uses a Neo4j Knowledge Graph database and a MAS for semi-automated IT-GS analysis based on the LangGraph~\cite{langgraph} framework. To build a solid base for all agents to be evaluated on, we use GPT-4.1~\cite{openai_models} for the KG construction and the initial transformation of the RecPlast documents into the graph structure. 
The audit-evaluation of the MAS is done with GPT-4o mini, GPT-4.1, GPT-5 mini, gpt-oss-120b~\cite{openai_models}, and Claude 4.5. Haiku~\cite{claude_4_5_haiku}. The operational agents use the Reasoning and Acting (ReAct) decision-making pattern~\cite{react} except GPT-5 mini (ReAct part removed), have tool-calls available to retrieve knowledge from the pre-computed KG and a SQL database to persist knowledge. 

\subsection{Knowledge Graph - Construction and Implementation}
The KG is generated by graph extraction using the LLMGraphTransformer~\cite{githubLLMGT} library using GPT-4.1 strictly for the ingestion phase of the IT-GS and RecPlast data. GPT 4.1 orchestrates the extraction of entities and relationships from the source documents into a searchable HybridRAG (Neo4j database). By using an affordable model with high performance, we create a robust, structural foundation that serves as a "guardrail" for the subsequent agents.
According to~\cite{graphtransformer}, LLMGraphTransformer shows significantly higher recall and better detection of functional relationships (e.g. causalities) compared to rule-based methods and was therefore selected.
Our solution implements a hybrid search within the Neo4j database, which combines semantic vector search with lexical full-text search. 
The Semantic Search is performed via text embeddings using vector indexes, and the full-text search via keyword indexes. Combining both methods for optimal retrieval quality yields the hybrid search approach.
This combination was selected to take advantage of the complementary strengths of both approaches. While vector search (text-embedding-3-small model~\cite{openaiembedding} captures semantic similarities and is robust against rephrasing, full-text search enables precise keyword matches, which is essential for technical terminology and requirement IDs in compliance documents. The results of both retrieval strategies are aggregated by a hybrid scoring mechanism in the retrieval pipeline, which combines cosine‑based vector similarity scores from Neo4j's vector index with BM25‑style relevance scores from Neo4j's full‑text index.

\subsection{Multi-Agent System Architecture and Agents}
Once the graph is established, the audit is performed step-by-step by our MAS. The MAS contains five agents: the \textit{Document Analysis agent}, the \textit{SA agent}, the \textit{PNs agent}, the \textit{Modeling agent}, and the \textit{IT-GS Check agent}. KG retrieval tools are bound and available to all agents. To evaluate the viability of this approach for resource-constrained SMEs, we vary the underlying LLM of the agents. We define GPT-4o mini as our baseline, representing a cost-optimized, high-speed option and compare it against GPT-5 mini and Claude 4.5 Haiku (efficient reasoning), GPT-OSS-120B (open weights for data sovereignty), and GPT-4.1 (to establish the upper performance bound).

\textbf{Structural Analysis Agent -}
The SA agent acts as a structured entity extractor, retrieving candidate nodes from the KG. To minimize the risk of omissions, it performs seven agent calls to extract possible entities, one for each BSI asset category, e.g., "Extract all \textit{applications} within the defined scope".
Unlike standard RAG approaches, the SA agent employs a \textit{Hypothesis-Verification Loop} for dependencies. First, it identifies implicit dependencies in the component descriptions (e.g., "uses central file server for data storage" implies a directed edge ("application", "uses", "central file server for data storage")). Then, it validates these hypothesized connections against the deterministic KG. If the target component (e.g., the specific file server) exists in the KG, the edge is confirmed. This rigorous check reduces the hallucination of non-existent dependencies and ensures a fully connected graph topology.

\textbf{Protection Needs Assessment Agent -}
The agent executes the two-stage reasoning pipeline to determine the CIA requirements. The PNs of the BPs are estimated by the agent (LLM-driven qualitative analysis) while the inheritance of the PNs (Maximum Principle, Cumulative Effect and Distribution Effect) are deterministic algorithms (algorithmic execution).
\begin{enumerate}
    \item \textbf{LLM-driven Qualitative Analysis:} The agent performs a damage analysis for each BP by querying the KG for damage scenarios and maps them to the BSI's CIA categories as they require semantic understanding of real-world consequences (e.g., interpreting "loss of reputation" or "GDPR violation"). This is the only LLM powered step in this pipeline.

    \item \textbf{Deterministic Propagation:} Once the initial CIA values are estimated, the propagation of PNs follows strict deterministic inheritance algorithms for the remaining three phases:

    \textbf{- Maximum Principle Analysis:} The agent traverses the dependency table. It applies the "Maximum Principle" by inheriting the highest protection need from dependent BPs down to all depended components both downward and upward through dependency chains. Conversely, it checks for "upward" consistency to ensure supporting assets do not undercut the security level of dependent processes.

    \textbf{- Cumulative Effect Analysis:} To detect hidden risks, the agent aggregates all components dependent on a single resource (e.g., a virtualization host). It evaluates whether the simultaneous failure of these components exceeds the impact of individual failures, potentially raising the PNs.

    \textbf{- Distribution Effect Analysis:} The Distribution Effect is a downward correction that addresses situations where an IT system serves highly critical components but only processes their non-critical parts, e.g., a server storing encrypted data without decryption keys. The Distribution Effect can only perform reductions, never increments. The agent checks if the initial protection level (from maximum principle/cumulative) is overestimated due to factors like partial data processing, pseudonymization, test data, or redundancies that mitigate potential damage.
\end{enumerate}

For components that go through all four principles, the meta rule applies. Should the principles lead to different results with regard to the final protection requirements, the meta rule selects the maximum from all four principles, thus ensuring that the most conservative approach is applied.

\textbf{Modeling Agent -}
The Modeling agent maps generic BSI modules to the infrastructure components according to the BSI methodology. It does so by querying the HybridRAG to retrieve the top-3 most suitable modules for each component. This threshold was derived from an IT-GS auditor feedback, indicating that while technically possible to map more modules, exceeding this threshold typically results in over-specification without significant security gains, while increasing the complexity for the human auditor verifying the results.
The automation allows for easy re-iteration. If the component description changes, e.g., the web server suddenly uses PostgreSQL instead of MySQL, the modeling can be rerun without requiring an auditor to repeat the entire analysis.

\textbf{IT-Grundschutz Check Agent -} 
The agent iterates through all applicable requirements defined by the selected modules and evaluates their degree of realization. It does so by querying the HybridRAG against the company's documents.   
For each requirement, the agent retrieves relevant context chunks and determines the degree of implementation ("Yes", "No", "Partially", "Dispensable"). Then it generates a citation-backed justification for each decision by explicitly linking the relevant document passage to the specific requirement.
Finally, the agent aggregates these individual verifications into a structured compliance report, which serves as the formal audit report required for the ISO 27001 certificate based on IT-GS.
\subsection{Data Set - RecPlast GmbH}\label{recplast_sec}
To evaluate the proposed MAS, we utilize the "RecPlast GmbH" case study developed by the BSI~\cite{recplast}. RecPlast GmbH is a fictitious manufacturing company used as a guiding example for the depth and scope of documentation required for a IT-GS certification~\cite{recplast}. The data set serves as a verifiable ground truth, as it allows for a holistic evaluation of our architecture. The data set consists of: 
\begin{itemize}
    \item \textbf{Organizational Context:} Workforce structure, locations, and Business Processes (BPs).
    \item \textbf{Strategic Documentation:} Mandatory policies and concepts, including information security guidelines, risk analysis policy, and record management policy.
    \item \textbf{Structural Analysis:} A complete infrastructure inventory of 134 entities (32 applications, 18 BPs, 3 ICS systems, 47 IT systems, 5 IoT systems, 8 communication connections, and 20 Rooms). 
    The data set DA lists 881 dependencies between the 134 different entities with six different relationship types including "Required for" (345), "Required" (345), "Includes" (93), "Located in" (92), "VM host for" (2), and "Virtualized on" (2) defining the topology for PNs inheritance.
    \item \textbf{Protection Needs Assessment:} A fully executed PNA defining CIA ratings for all 134 entities. For each entity, the PNs (Normal/High/Very High) are distributed as follows; Confidentiality (13/102/18), Integrity (35/94/4), and Availability (60/34/39).
    \item \textbf{Modeling:} A complete Modeling, where the infrastructure components are mapped onto specific IT-GS modules.The data set lists 134 components in the SA, of which 84 have modules assigned. Those 84 components are assigned to 137 modules from the IT-GS compendium. A total of 49 different modules are used. 
    \item \textbf{IT-GS Check:} A completed IT-GS Check and a finalized risk treatment plan covering 3497 individual requirements, each of them being assigned to one of four levels of fulfillment; "Yes" (2854), "No" (97), "Partially" (166), or "Dispensable" (380) providing a granular basis for precision and recall testing.
\end{itemize}
Unlike proprietary real-world data (often obfuscated due to confidentiality), RecPlast provides a transparent, and publicly available data set for the entire documentation stack. This enables the rigorous evaluation of quantitative metrics, such as recall and precision, verifying our MAS against a deterministic and expert-generated reference data set.
\section{Results \& Discussion}
The RecPlast data set includes intermediate results for each step of the certification process. To accurately quantify the performance of each specific agent without interference from subsequent error propagation, a step-wise evaluation is performed.
Performance is measured using standard information retrieval metrics: Precision, Recall, and F1-score~\cite{augementedSOC}. For entity extraction and matching, Cosine Similarity (CS) is utilized over vector embeddings. Although embeddings are susceptible to Representation Degeneration, empirical studies on large-scale content analysis in comparable domains (formal regulatory documentation) indicate that optimal thresholds for CS typically converge around 0.6~\cite{repdeg}. Specifically, ~\cite{CScore} identified optimal thresholds in the range of 0.57 to 0.63 when analyzing corporate disclosure documents for current embedding models (GPT text-embedding-3-small, Gemini text-embedding-004). Based on this empirical evidence, a threshold of 0.6 is set for this evaluation to ensure robust matching of entities.

\textbf{Structural Analysis -}
\begin{table}[htbp]
  \centering
  \caption{Entity Extraction (CS = 0.6)\\Matched: (extracted ground truth entries / all extracted entities)}
  \label{tab:sa}
  \begin{tabular}{l|c|c|c|c}
    \textbf{Model} & \textbf{Matched} & \textbf{Precision} & \textbf{Recall} & \textbf{F1-score} \\
    \hline
    GPT-4o mini & 58/98 & \textbf{59.2\%} & 43.9\% & 50.4\% \\ 
    GPT-4.1 & 68/174 & 40.7\% & 51.5\% & 45.5\% \\
    GPT-5 mini & 70/177 & 39.8\% & 53.4\% & 45.6\% \\
    Claude 4.5 Haiku & 75/188 & 41.4\% & \textbf{57.3\%} & 48.1\% \\
    GPT-OSS-120B & 74/150 & 49.7\% & 55.2\% & \textbf{52.3\%} \\
 
  \end{tabular}
\end{table}
The SA shows the results of entity extraction, as illustrated in Table~\ref{tab:sa}. GPT-OSS-120B achieves the highest balance between precision (49.7\%) and recall (55.2\%) with an F1-score of 52.3\%, identifying 74 of the 134 entities. GPT-4o mini shows the highest precision (59.2\%), but the lowest recall (43.9\%) with only 58 correctly recognized entities. Claude 4.5 Haiku achieves the highest recall (57.3\%) and the highest number of true positives (75), but tends to over-extract with 188 entities found, 40 more than in the data set. GPT-4.1 and GPT-5 mini show comparable F1-scores (45.5\% and 45.6\%) with a similar level of over-extraction.
This leads to the question whether the additional extracted entities are hallucinations or not. A vector search is carried out to determine whether the extracted entities could be extracted from the data or were invented. In all cases, the extracted entities could be traced back within the company documents and were not hallucinated.
The results indicate a precision-recall trade-off among agents. While over-extraction implies additional validation effort, in an audit context, Recall is strictly prioritized over Precision. Missing an asset creates a critical blind spot in the security concept, whereas filtering out a non-existent device is merely a cleanup task.
This assessment is confirmed by an IT-GS auditor interviewed, who stated that he considers the extraction of additional components to be beneficial, as particularly IoT devices are often overlooked, e.g., internet-connected vacuum cleaners posing a relevant threat but are rarely found in SAs. Further, he pointed out that it is an acceptable trade-off to go through proposed entities compared to the amount of time required for manual extraction from documents. Last, it should be noted that GPT-OSS-120B, as an open-source model, outperforms proprietary alternatives in terms of F1-score, hence, offering a data protection-compliant option for companies with increased data sovereignty requirements.  

For the DA evaluation, only exact matches of components and the type of relationships mentioned above were accepted.
%
 \begin{table}[htbp]
      \centering
      \caption{Dependency Analysis (CS = 0.6)}
      \label{tab:abhaengigkeiten-evaluation}
      \begin{tabular}{l|r|r|r|r}
      \textbf{Model} & \textbf{Matched} & \textbf{Precision} & \textbf{Recall} & \textbf{F1-Score} \\
      \hline
      GPT-4o mini      & 433/455   & 69.1\% & 35.7\% & 47.1\% \\
      GPT-4.1          & 857/1578  & 38.2\% & 68.4\% & 49.0\% \\
      GPT-5 mini       & 832/1559  & 34.7\% & 61.5\% & 44.4\% \\
      Claude 4.5 Haiku & 854/1416  & 43.6\% & 70.1\% & \textbf{53.8\%} \\
      gpt-oss-120b     & 485/614   & 51.7\% & 36.0\% & 42.5\% \\
      \end{tabular}
  \end{table}
DA evaluation reveals significant performance differences, as illustrated in Table~\ref{tab:abhaengigkeiten-evaluation}. Claude 4.5 Haiku achieves the highest F1-score of 53.8\% and identifies 70.1\% of all data set dependencies with a precision of 43.6\%. GPT-4.1 follows with an F1-score of 49.0\% and a similarly high recall of 68.4\%, but produces more false positives. GPT-4o mini shows the opposite behavior, with a precision of 69.1\% making it the most accurate LLM, but only finding 35.7\% of the actual dependencies. The lowest scores are achieved by GPT-5 mini and gpt-oss-120b with F1-scores of 44.4\% and 42.5\%, respectively.
Noteworthy is the high proportion of partial matches in all LLMs where the components were correctly recognized, but the relationship type did not match. In GPT-5 mini, the number of partial matches (581) even exceeds that of exact matches (251), indicating systematic weaknesses in type classification. The LLMs can be divided into two groups: GPT-4o mini and gpt-oss-120b extract fewer but more precise dependencies, while GPT-4.1, GPT-5 mini, and Claude 4.5 Haiku achieve higher recall at the expense of precision.
With F1-scores peaking at only 53.8\% (Claude 4.5 Haiku), the results demonstrate that current LLMs struggle to reliably reconstruct the precise dependency topology required by the IT-GS. A detailed error analysis reveals that LLMs frequently hallucinate plausible dependencies based on general IT training data (e.g., assuming a web server always connects to a specific database type) rather than strictly extracting evidence-based connections from the company documents.
The high percentage of false positives indicates that LLMs have difficulty distinguishing between plausible and actual dependencies. This limitation has critical implications for the safety of the audit. These results empirically validate the strict necessity of a "Human-in-the-Loop" specifically at this step. The MAS is capable of proposing a candidate graph of dependencies to speed up the process, but the auditor must act as a corrective instance to validate the integrity.

\textbf{Protection Needs Assessment -}    
The distribution of the CIA goals for each entity, mentioned in ~\ref{recplast_sec}, shows characteristic variations between the dimensions, as for confidentiality, "High" dominates with 102 entries (78\%), while "Normal" and "Very High" occur rarely. Integrity follows a similar pattern with 94 high entries (72\%), but contains more normal classifications (35) and only 4 "Very High" cases. Availability shows the most balanced distribution, as "Normal" is the most common with 60 entries (46\%), followed by "Very High" (39) and "High" (34).
\begin{table}[htbp]
\centering
\caption{Classification accuracy for Protection Needs\\Models: 4o-m = GPT-4o mini, 4.1 = GPT-4.1, 5-m = GPT-5 mini, Haiku = Claude 4.5 Haiku, OSS = gpt-oss-120b.}
\label{tab:schutzbedarf-gesamt}
\begin{tabular}{l|ccc|ccc|ccc}
& \multicolumn{3}{c|}{\textbf{Confidentiality}} & \multicolumn{3}{c|}{\textbf{Integrity}} & \multicolumn{3}{c}{\textbf{Availability}} \\
\hline
\textbf{Model} & N & H & VH & N & H & VH & N & H & VH \\
4o-m    & 0.0 &  6.1 & \textbf{88.2} & 0.0 &  3.8 & \textbf{100.0} & 0.0 & 0.0 & 97.1 \\
4.1        & 0.0 & \textbf{95.1} & 12.5 & 4.5 & \textbf{93.7} &   0.0 & 4.1 & 8.7 & 97.0 \\
5-m     & 0.0 & \textbf{96.3} & 12.5 & 0.0 & 41.8 &  25.0 & 0.0 & 4.3 & 97.0 \\
Haiku      & 0.0 & 37.8 & 25.0 & 0.0 & 41.8 &   0.0 & 0.0 & 8.7 & \textbf{100.0} \\
OSS   & 0.0 & 37.8 & 25.0 & 0.0 & 41.8 &   0.0 & 0.0 & 8.7 & \textbf{100.0} \\
\end{tabular}
\end{table}
The evaluation of the PNs classification is illustrated in Table~\ref{tab:schutzbedarf-gesamt}. 
\textbf{Confidentiality:} GPT-4o mini classifies almost all entities as "Very High" (VH), achieving a recall of 88.2\% for this category, but producing 2 critical under-protection cases. GPT-4.1 and GPT-5 mini, recognize the "High" (H) category with over 95\% accuracy, but often downgrade "Very High" entities, resulting in 14 and 15 under-protected cases. No model correctly recognizes "Normal" (N).
\textbf{Integrity:} GPT-4o mini achieves a perfect recall of 100\% for "Very High", as it classifies almost all entities in this category. There is a single case of under-protection. GPT-4.1 shows a contrasting behavior, as it classifies "High" with 93.7\% accuracy but fails to correctly identify any "Very High" entities. GPT-4.1 is the only model that identified "Normal" in isolated cases (4.5\%). Claude 4.5 Haiku and gpt-oss-120b show identical results with moderate high detection (41.8\%) and no "Very High" detection.
\textbf{Availability:} All LLMs show a strong tendency to classify as "Very High" with recall values between 97\% and 100\%. Claude 4.5 Haiku and gpt-oss-120b achieve 100\% for "Very High" with only one under-protection case. The detection of "High" is low across all LLMs (0-8.7\%), and "Normal" is almost never detected, despite the data set containing 60 entries, representing the most "Normal" classifications in this dimension. GPT-4.1 is again the only model with minimal "Normal" detection (4.1\%).      
Another result is the inability of all LLMs to correctly classify the category "Normal". This causes systematic overestimation, while overprotection poses no direct security risk, it increases implementation effort and costs due to insufficient organization-specific context in the training data to distinguish critical from non-critical systems.
GPT-4o mini reduces under-protection cases (only 4 in total) by strongly classifying them as "Very High", but only achieves 6\% recall for "High". GPT-4.1, on the other hand, detects "High" with over 93\% accuracy, but produces 23 under-protection cases.
Particularly critical is the inability to recognize "Very High" in GPT-4.1 and GPT-5 mini. These LLMs often downgrade highly critical entities to "High", which poses a potential threat to security.                                   
In terms of confidentiality, GPT-4.1/GPT-5 mini differentiate most effectively between "High" and other categories (95\%), but lacks accuracy for "Very High". Integrity is the weakest dimension, with inconsistent results. Only GPT-4o mini correctly identifies "Very High" (100\%), but through over-classification. All LLMs show a tendency to overestimate availability as "Very High" (97–100\%), although 46\% of the data set entries classify as "Normal".
The evaluation indicates that the LLMs are capable of classifying PNs, but with a clear tendency to overestimate them. However, this also corresponds to the human approach, as confirmed by the auditor interviewed, who stated that in case of doubt, it is preferable to take a more conservative approach.

\textbf{Modeling and Module Assignment -}
The modules used most frequently are SYS.1.1 (general server), SYS.3.1 (laptops), SYS.1.2.2 (Windows server), INF.7 (office workstation), SYS.2.1 (general client), and SYS.2.2.3 (Windows clients).
\begin{table}[htbp]
  \centering
  \caption{Modeling: Mapping Components to Modules}
  \label{tab:modellierung-evaluation}
  \begin{tabular}{l|r|r|r|r}
  \textbf{Model} & \textbf{Precision} & \textbf{Recall} & \textbf{F1-score} & \textbf{Coverage} \\
  \hline
  GPT-4o mini      & 25.7\% & 32.1\% & 28.6\% & 40.5\% \\
  GPT-4.1          & 36.0\% & 19.7\% & 25.5\% & 19.0\% \\
  GPT-5 mini       & 26.5\% & 73.0\% & 38.8\% & 96.4\% \\
  Claude 4.5 Haiku & 38.6\% & 74.5\% & \textbf{50.9\%} & 96.4\% \\
  gpt-oss-120b     & 29.2\% & 73.0\% & 41.7\% & 95.2\% \\
  \end{tabular}
  \end{table}
Claude 4.5 Haiku achieves the highest F1-score of 50.9\%. The model finds 102 of the 137 correct mappings (recall 74.5\%), but also generates 162 incorrect mappings (precision 38.6\%). Component coverage is 96.4\%, meaning that modules were suggested for 81 of the 84 components. gpt-oss-120b follows with an F1-score of 41.7\% and achieves a high recall of 73.0\% (100 correct assignments), but with 243 false positives, it has a significantly lower precision of 29.2\%. GPT-5 mini follows with an F1-score of 38.8\%. The recall is identical to gpt-oss-120b at 73.0\%, but the model produces even more false positives with 278 incorrect assignments, resulting in a precision of 26.5\%, as illustrated in Table~\ref{tab:modellierung-evaluation}.
The lowest-scoring LLMs are GPT-4o mini and GPT-4.1. GPT-4o mini achieves an F1-score of 28.6\% with a component coverage of 40.5\%, which shows that half of the components have no assignments. GPT-4.1 performs worse with an F1-score of 25.5\%. Although precision is comparatively high (36.0\%), recall is only 19.7\%, since only 16 of the 84 components were processed at all. 
Component coverage reveals a clear difference between the LLMs. GPT-4o mini and GPT-4.1 do not perform modeling on the majority target objects; the newer LLMs GPT-5 mini, Claude 4.5 Haiku, and gpt-oss-120b achieve almost complete coverage (> 95\%). Indicating that these LLMs handle the task more robustly and attempt to assign even less obvious target objects.
The analysis of errors reveals certain patterns, such as process-oriented modules like CON, OPS, or ISMS often being incorrectly assigned to applications. In addition, LLMs frequently confuse related modules, such as APP.2.2 instead of APP.2.1 (Active Directory). However, operating system-specific modules such as SYS.2.2.3, cloud-related modules such as OPS.2.2, and the INF.7 module (office workstation) for office spaces are often missing.

\textbf{IT-Grundschutz Check -}
\begin{table}[htbp]
\centering
\caption{Predictions by category}
\label{tab:predictions-per-category}
\begin{tabular}{l|c|c|c|c}
\textbf{Model} & \textbf{Yes} & \textbf{No} & \textbf{Partially} & \textbf{Dispensable}  \\
\hline
GPT-4.1          & 1212 & 1634 & 389 & 262  \\
GPT-4o mini      & 1125 & 1477 & 354 & 541  \\
gpt-oss-120b     & 1247 & 1457 & 384 & 409   \\
Claude 4.5 Haiku & 249  & 2143 & 709 & 396   \\
\end{tabular}
\end{table}
The evaluation looked at five LLMs in their ability to classify the degree of implementation of IT-GS requirements, as shown in Table~\ref{tab:predictions-per-category}.  GPT-4.1 achieved the highest accuracy at 32\%, followed by GPT-4o mini (21\%) and gpt-oss-120b (19\%). 
The evaluation reveals a massive limitation of the proposed architecture as the error analysis uncovers a significant discrepancy in the distribution. While the data set reflects a highly compliant organization, LLMs cannot verify this, resulting in a high rate of false negatives. 
This systematic blind spot shows that LLMs can retrieve relevant passages but struggle to reason whether content sufficiently verifies a given fulfillment level—they cannot distinguish mentioning a security concept from proving its implementation. These results confirm that LLMs cannot serve as autonomous compliance judges. Their probabilistic nature conflicts with the deterministic verdict required by the BSI, limiting their role to a heuristic search engine proposing evidence candidates for mandatory human adjudication.
\section{Limitations \& Future Work}
A key limitation of the proposed MAS architecture is the lack of retrospective error correction. Our evaluation shows that errors in the DA have a direct impact on the PNA and compound throughout the audit workflow. Since the determination of PNs is based on strict inheritance rules, even a single missing edge in the graph topology causes subsequent calculations such as the "maximum principle" or the "cumulation effect" to become mathematically incorrect. In a compliance context, this severed link breaks the chain of inheritance, meaning high protection needs fail to propagate to supporting infrastructure, leaving critical assets effectively under-protected.
Secondly, the low accuracy of the IT-GS check (32\%) highlights a fundamental gap. Although agents successfully extract relevant components, they fail to reliably translate this semantic context into fulfillment judgments. The observed blind spot for evidence suggests that current LLMs are not yet capable of the precise logical deduction required to perform full audits. 
Finally, although the RecPlast dataset enables precise metric calculation, it represents a "best-case" scenario with standardized terminology. Real-world company data are unstructured, ambiguous, and fragmented. Given that state-of-the-art LLMs encounter difficulties with the explicit structures of RecPlast, it must be assumed that performance on noisy real-world data would be significantly lower without additional layers of pre-processing.
Future research will consequently shift the focus from full automation to interactive "Human-in-the-Loop" architectures. Instead of replacing the auditor, our plan is to implement iterative feedback loops in which the agent suggests graph topologies and compliance notes but requires confirmation from an auditor before proceeding to the next step. 
\section{Conclusion}
This paper provided the first empirical end-to-end evaluation of the HybridRAG-based MAS to support IT-GS audits. Through the expert-reviewed "RecPlast" dataset, it was possible to extend beyond purely qualitative assessments and quantitatively evaluate the performance of the MAS in a strict regulatory context. Our results reveal a fundamental problem with using LLMs for compliance tasks. Although the architecture delivers excellent results in semantic information extraction, it currently lacks the logical precision required for deterministic certification.
The empirical data show that modern LLMs serve effectively as heuristic tools for the SA and modeling phases. The high recall in entity extraction demonstrates that the approach can significantly reduce the manual effort required to capture and assign infrastructure components. However, the low F1-scores in DA and the insufficient accuracy in the IT-GS check illustrate that the probabilistic nature of LLMs conflicts with the strict logic of the SA inheritance rules. A single hallucinated/forgotten dependency interrupts the entire chain of PNA, making a fully automated judgment impossible.
Our conclusion is that the role of AI in safety-critical audits needs to be redefined from "autonomous auditor" towards "intelligent pre-processor". The proposed MAS architecture is valuable because it identifies potential gaps. Future research should  prioritize a "Human-in-the-Loop" approach, in which the agent refines the KG based on explicit feedback from the auditor, rather than deciding autonomously on final compliance decisions.


\begin{thebibliography}{00}
%
\bibitem{NIS2} European Parliament and Council of the European Union, ``Directive (EU) 2022/2555 of the European Parliament and of the Council of 14 December 2022 on measures for a high common level of cybersecurity across the Union, amending Regulation (EU) No 910/2014 and Directive (EU) 2018/1972, and repealing Directive (EU) 2016/1148 (NIS 2 Directive),'' Official Journal of the European Union, vol. L333, pp. 80--152, Dec. 2022.
%
\bibitem{iso_27001} ISO/IEC 27001:2022. Information security, cybersecurity and privacy protection — Information security management systems — Requirements. Edition 3, 2022. Available: \url{https://www.iso.org/standard/27001}. (Accessed: 2025-12-21)
%
\bibitem{grundschutz_standards} BSI, ``BSI-Standards,'' Bundesamt für Sicherheit in der Informationstechnik. Available: \url{https://www.bsi.bund.de/dok/6603458}. (Accessed: 2025-12-21)
%
\bibitem{BSI_NIS2_2025} Bundesamt für Sicherheit in der Informationstechnik (BSI). \emph{IT-Sicherheitsrecht: NIS-2-Regierungsentwurf ist ein großer Schritt auf dem Weg zur Cybernation}. Pressemitteilung, 30. July 2025. Available: \url{https://www.bsi.bund.de/DE/Service-Navi/Presse/Pressemitteilungen/Presse2025/250730_NIS-2-Regierungsentwurf.html}.
(Accessed: 2025-12-21)
%
\bibitem{enisa} European Union Agency for Cybersecurity (ENISA), ``Navigating cybersecurity investments in the time of NIS 2,'' ENISA, Jul. 2023. Available: \url{https://www.enisa.europa.eu/news/navigating-cybersecurity-investments-in-the-time-of-nis-2}. (Accessed: 2025-12-21)
%
\bibitem{former_paper} L. Muth and M. Margraf, ``An Approach for a Supporting Multi-LLM System for Automated Certification Based on the German IT-Grundschutz,'' 2025 IEEE International Conference on Cyber Security and Resilience (CSR), Chania, Crete, Greece, 2025, pp. 482-489. Available: \url{https://doi.org/10.1109/CSR64739.2025.11130171}. 
%
\bibitem{hybridRAG} B. Sarmah et al., ``HybridRAG: Integrating Knowledge Graphs and Vector Retrieval Augmented Generation for Efficient Information Extraction,'' Proceedings of the 5th ACM International Conference on AI in Finance, 608--616, Nov. 2024. Available: \url{https://doi.org/10.1145/3677052.3698671}. 
%
\bibitem{graphRAG_OG} D. Edge et al., ``From Local to Global: A Graph RAG Approach to Query-Focused Summarization,'' arXiv preprint arXiv:2404.16130, Apr. 2024. Available: \url{https://doi.org/10.48550/arXiv.2404.16130}. 
%
\bibitem{recplast} Bundesamt für Sicherheit in der Informationstechnik, ``IT-Grundschutz-Kompendium: Hilfsmittel und Anwenderbeiträge,'' BSI, 2023. Available: \url{https://www.bsi.bund.de/DE/Themen/Unternehmen-und-Organisationen/Standards-und-Zertifizierung/IT-Grundschutz/Hilfsmittel_und_Anwenderbeitraege/Hilfsmittel_vom_BSI/Recplast/recplast_node.html}. (Accessed: 2025-12-21)
%
\bibitem{grundschutz_kompendium} Bundesamt für Sicherheit in der Informationstechnik, ``IT-Grundschutz-Kompendium,'' Edition 2023. Available: \url{https://www.bsi.bund.de/SharedDocs/Downloads/DE/BSI/Grundschutz/IT-GS-Kompendium/IT_Grundschutz_Kompendium_Edition2023.pdf}. (Accessed: 2025-12-21)
%
\bibitem{old2} S. Fenz and T. Neubauer, ``Ontology-based information security compliance determination and control selection on the example of ISO 27002,'' Information \& Computer Security, Vol. 26, Nov. 2018. Available: \url{https://doi.org/10.1108/ICS-02-2018-0020}
%
\bibitem{pilot_iso_27001} A. Salman, Y. Alsiyat, S. Creese, and M. Goldsmith, ``Work in Progress: Leveraging Large Language Models for Cybersecurity Compliance: A Pilot Study in ISO 27001 Audit Planning,'' 2025 IEEE European Symposium on Security and Privacy Workshops, pp. 351--359, Jun. 2025. Available: \url{https://doi.org/10.1109/EuroSPW67616.2025.00046}. (
%
\bibitem{CCKG4HLRoV} J. Vizcarra, Y. Gempei, Y. Wang, T. Isohara, and M. Kurokawa, ``Constructing Cybersecurity Knowledge Graphs for Hybrid LLM–Graph Reasoning on Vulnerabilities,'' ISWC 2025 Companion Volume, Nov. 2025. Available: \url{https://ceur-ws.org/Vol-4085/paper35.pdf}
%
\bibitem{CyKGRAG} K. Kurniawan, E Kiesling, and A. Ekelhart, ``CyKG-RAG: Towards knowledge-graph enhanced retrieval augmented generation for cybersecurity,'' RAGE-KG 2024 Workshop at ISWC 2024, Nov. 2024.  Available: \url{https://ceur-ws.org/Vol-3950/paper1.pdf}. (Accessed: 2025-12-21)
%
\bibitem{langgraph} LangGraph AI. Available: \url{https://github.com/langchain-ai/langgraph}. (Accessed: 2025-12-21)
%
\bibitem{openai_models} OpenAI, GPT-4o mini, GPT-4.1, GPT-5 mini, and GPT-OSS-120B. Available: \url{https://platform.openai.com/docs/models/}. (Accessed: 2025-12-21)
%
\bibitem{react} S. Yao, J. Zhao, D. Yu, N. Du, and I. Shafran, ``React: Synergizing reasoning and acting in language models,'' ICLR 2023, Feb. 2023.  Available: \url{https://arxiv.org/abs/2210.03629}. 
%
\bibitem{claude_4_5_haiku} Anthropic, Claude Haiku 4.5. Available: \url{https://www.anthropic.com/news/claude-haiku-4-5}. (Accessed: 2025-12-21)
%
\bibitem{githubLLMGT} LLMGraphTransformer. Available: \url{https://github.com/dhiaaeddine16/LLMGraphTransformer}. (Accessed: 2025-11-2)
%
\bibitem{graphtransformer} V. E. Yamamoto et al., ``Exploring LLM To Extract Knowledge Graph From Academic Abstracts,'' ISWC 2025 Companion Volume, Nov. 2025. Available: \url{https://ceur-ws.org/Vol-4085/paper49.pdf}. 
%
\bibitem{openaiembedding} OpenAI, text-embedding-3-small Model. Available: \url{https://platform.openai.com/docs/models/text-embedding-3-small}. (Accessed: 2025-12-21)
%
\bibitem{augementedSOC} S. Srinivas et al., ``AI-Augmented SOC: A Survey of LLMs and Agents for Security Automation,'' Journal of Cybersecurity and Privacy, Vol. 5, Article Nr. 95, Sep. 2025. Available: \url{https://doi.org/10.3390/jcp5040095}. 
%
\bibitem{repdeg} Z. Zhang et al., ``Revisiting Representation Degeneration Problem in Language Modeling,'' In Findings of the Association for Computational Linguistics: EMNLP 2020, pp. 518–-527, Nov. 2020. Available: \url{https://doi.org/10.18653/v1/2020.findings-emnlp.46}.
%
\bibitem{CScore} M. Kamat, J. Jagasia, A. Vaidya, and O. Surve, ``Embedding-Based decision support framework for large-scale content analysis'', Knowledge-Based Systems, Volume 332, Nov. 2025. Available: \url{https://doi.org/10.1016/j.knosys.2025.114926}. (Accessed: 2025-12-21)
%
 

 
 
 
 
 

\end{thebibliography}
\end{document}